\documentclass[conference]{IEEEtran}
\IEEEoverridecommandlockouts
\usepackage{cite}
\usepackage{amsmath,amssymb,amsfonts}
\usepackage{algorithmic}
\usepackage{graphicx}
\usepackage{textcomp}
\usepackage{xcolor}

\newcommand{\lnorm}{\left\Vert}
\newcommand{\rnorm}{\right\Vert}

\usepackage{graphicx}
\usepackage{acronym}
\newacro{ODE}{ordinary differential equation}
\newacroindefinite{ODE}{an}{a}
\newacroindefinite{EQL}{an}{a}
\newacro{PINN}{Physics-Informed Neural Network}
\newacro{CODE}{continuous ordinary differential equation}
\newacro{EKBF}{extended Kalman-Bucy-Filter}
\newacroindefinite{EKBF}{an}{a}
\newacro{EQL}{Equation Learner}
\newacro{KBINN}{Kalman-Bucy-Informed Neural Network}
\newacro{INN}{Integrated Neural Network}
\newacroindefinite{INN}{an}{a}
\newacro{RL}{Reinforcement Learning}
\newacro{MDP}{Markov Decision Process}
\newacro{MLP}{multi-layer perceptron}
\newacroindefinite{mlp}{an}{a}
\newacro{SINDy}{Sparse Identification of Nonlinear Dynamics}
\newacro{MPC}{model-based predictive controller}
\newacroindefinite{MPC}{an}{a}
\newacro{LQR}{Linear Quadratic Regulator}

\usepackage{amsfonts}
\usepackage{amsmath}
\usepackage{bm}
\usepackage{siunitx}

\usepackage{tikz}
\usepackage{pgfplots}
\pgfplotsset{compat=1.8}
\usepgfplotslibrary{statistics}
\usepackage{caption}
\usepackage{subcaption}
\usepgfplotslibrary{groupplots}
\newcommand\mydots{\hbox to 1em{.\hss.\hss.}}
\usepgfplotslibrary{fillbetween}

\usepackage{listings}

\usepackage[utf8]{inputenc} 
\usepackage[T1]{fontenc}    
\usepackage{url}            
\usepackage{booktabs}       
\usepackage{amsfonts}       
\usepackage{nicefrac}       
\definecolor{mygreen}{RGB}{0,150,0}

\usepackage{cite}
\usepackage{amsmath,amssymb,amsfonts}
\usepackage{algorithmic}
\usepackage{graphicx}
\usepackage{algorithm,algorithmic}
\usepackage{textcomp}
\usepackage[hidelinks]{hyperref}

\usepackage{pgfplotstable}
    \usepgfplotslibrary{statistics}
    \usepgfplotslibrary{colorbrewer}
    \pgfplotsset{
        compat=1.16,
        cycle list/Set3-4,
        /pgf/declare function={
            Floor(\x) = round(\x-0.49);
        },
    }

\usepackage{placeins}
\usepackage{dblfloatfix}
\def\BibTeX{{\rm B\kern-.05em{\sc i\kern-.025em b}\kern-.08em
    T\kern-.1667em\lower.7ex\hbox{E}\kern-.125emX}}
\begin{document}

\title{Identifying Ordinary Differential Equations for Data-efficient Model-based Reinforcement Learning\\
\thanks{This work was supported by the Baden-Wuerttemberg Ministry for Economic Affairs, Labour and
Tourism (project KI-Fortschrittszentrum ``Lernende Syteme und Kognitive Robotik'', grant no. 036-140100).}
}

\author{\IEEEauthorblockN{Tobias Nagel}
\IEEEauthorblockA{\textit{Cyber Cognitive Intelligence} \\
\textit{Fraunhofer Institute for Manufacturing}\\
\textit{Engineering and Automation} \\
70569 Stuttgart, Germany \\
tobias.nagel@ipa.fraunhofer.de}
\and
\IEEEauthorblockN{Marco F. Huber}
\IEEEauthorblockA{\textit{Cyber Cognitive Intelligence} \\
\textit{Fraunhofer Institute for Manufacturing}\\
\textit{Engineering and Automation} \\
70569 Stuttgart, Germany \\
marco.huber@ieee.org}
}

\author{\IEEEauthorblockN{Tobias Nagel\IEEEauthorrefmark{1} and Marco F. Huber\IEEEauthorrefmark{1}\IEEEauthorrefmark{2}}
\IEEEauthorblockA{\IEEEauthorrefmark{1}\textit{Fraunhofer Institute for Manufacturing Engineering and Automation IPA, 70569 Stuttgart, Germany} \\  
tobias.nagel@ipa.fraunhofer.de}
\IEEEauthorblockA{\IEEEauthorrefmark{2}\textit{Institute of Industrial Manufacturing and Management IFF, University of Stuttgart, 70569 Stuttgart, Germany} \\  
marco.huber@ieee.org}
}

\maketitle

\begin{abstract}
The identification of a mathematical dynamics model is a crucial step in the designing process of a controller. However, it is often very difficult to identify the system's governing equations, especially in complex environments that combine physical laws of different disciplines. In this paper, we present a new approach that allows identifying an ordinary differential equation by means of a physics-informed machine learning algorithm. Our method introduces a special neural network that allows exploiting prior human knowledge to a certain degree and extends it autonomously, so that the resulting differential equations describe the system as accurately as possible. We validate the method on a Duffing oscillator with simulation data and, additionally, on a cascaded tank example with real-world data. Subsequently, we use the developed algorithm in a model-based reinforcement learning framework by alternately identifying and controlling a system to a target state. We test the performance by swinging-up an inverted pendulum on a cart.
\end{abstract}


\section{Introduction}
\label{sec:introduction}
Identifying the characteristics and behavior of a system is a vital step for predicting and controlling its future states. In dynamic systems, the governing equation is often represented by means of an \ac{ODE} that contains derivatives with respect to time. However, in many cases it is very difficult to derive the equations that describe the underlying physical laws because large systems in particular exhibit complex behavior. In this paper, we present a new method that allows identifying a nonlinear \ac{ODE} from noisy input-output data, which is applicable in a model-based \ac{RL} framework. Our method allows human knowledge to be incorporated when available, making it a hybrid machine learning approach. 

In the following, we give a formal problem statement. 
The state-space representation of a continuous-time, nonlinear, dynamic, stochastic, and time-variant system of rank $n \in \mathbb{N}$ is given by
\begin{equation} \label{eq:ss_representation}
\begin{split}
    \bm{\dot{x}}(t) &= \bm{f}\left( \bm{x}(t), \bm{u}(t), \bm{w}(t),t\right),~ \bm{x}(t_0) = \bm{x}_0~, \\
    \bm{y}(t) &= \bm{g}\left( \bm{x}(t), \bm{u}(t), \bm{v}(t),t \right) ~,
\end{split}
\end{equation}
with the nonlinear \ac{ODE} $\bm{f}(\cdot)$ and the measurement function $\bm{g}(\cdot)$. $\bm{x}(t) \in \mathbb{R}^n$ denotes the state vector with an initial value $\bm{x}_0$. $\bm{u}(t) \in \mathbb{R}^p$ and $\bm{y}(t) \in \mathbb{R}^q$ denote the input and output signal with dimensions $p,q \in \mathbb{N}$, respectively. The vectors $\bm{w}(t)$ and $\bm{v}(t)$ denote white process noise and white measurement noise, respectively, which are both assumed to be zero-mean Gaussian with covariance matrices $\bm{Q}(t) \in \mathbb{R}^{n\times n}$ and $\bm{R}(t) \in \mathbb{R}^{q\times q}$, respectively. We assume that we can acquire noisy measurements $\overline{\bm{y}}(t_i)$ of the system at $N$ discrete time steps $t_i$ with $i = 1,\ldots,N$. 

Depending on the use case, we distinguish three situations: (i) $\bm{f}$ and $\bm{g}$ are known, apart from some real-valued parameters. (ii) $\bm{f}$ and $\bm{g}$ are partly known, which means that it is possible to model single equation elements, but not the system as a whole. (iii) $\bm{f}$ and $\bm{g}$ are unknown, apart from the system states and the system rank. The aim of our method is to identify $\bm{f}\left( \cdot \right)$ and $\bm{g}\left( \cdot \right)$ in such a way that
\begin{equation}
    J = \sum_{i=1}^{N} \left| \bm{y}(t_i) - \overline{\bm{y}}(t_i) \right|_\mathrm{F}
\end{equation}
is minimal, with $\left| \cdot \right|_\mathrm{F}$ being the Frobenius norm.

In this paper, we present a new method to identify \acp{ODE} that we call \emph{\ac{ODE}-Learner}. We use an \ac{EKBF} (cf. Sec.~\ref{sec:fundamentals_ekbf}), which is a continuous-time and nonlinear implementation of the famous Kalman filter, consisting of two \acp{ODE} for estimating the mean value and the covariance matrix of the system state. As $\bm f(\cdot)$ and $\bm g(\cdot)$ are assumed to be (partly) unknown, the \ac{EKBF} cannot be applied directly. Instead, both \acp{ODE} of the \ac{EKBF} are realized as neural networks in a \ac{PINN}-framework (cf. Sec.~\ref{sec:fundamentals_pinn}). The governing equation of the \ac{ODE} is learned  using two further neural networks in an \ac{EQL}-framework (cf. Sec.~\ref{sec:fundamentals_eql}). 

A field of application is model-based \ac{RL}, where an agent collects data in an environment and trains a model that allows predicting the future system behavior. The model is then used to calculate a reward-maximizing policy, which is applied to the agent. Thus, more data can be gathered and the model is re-trained to improve its accuracy until, finally, a target state is reached. This procedure is often more data-efficient than model-free alternatives~\cite{Donge.2023}. We show how the \ac{ODE}-Learner can be used within continuous-time model-based \ac{RL} to learn the dynamics model of the agent. The key contributions of our method comprise:
\begin{itemize}
    \item Identifying an \ac{ODE} from noisy data and not directly accessible system states by combining an \ac{EKBF} with \acp{PINN} and \acp{EQL}.
    \item Extending the \ac{EQL} framework to create more shallow networks.
    \item Allowing to incorporate human knowledge if available.
    \item Using the identified \ac{ODE} in a model-based \ac{RL} environment.
\end{itemize}
Please note that opposed to many existing methods, we do not simplify the identification process by fitting a discrete-time difference equation, but train in a fully continuous-time fashion, which features several advantages. Firstly, model building typically relies on physical equations that describe dynamics in continuous time. The continuous-time identification approach allows the direct integration of existing system knowledge into the training algorithm. Secondly, continuous-time identification eliminates the need for a pre-defined sampling rate. In contrast, discrete-time identification necessitates setting a sampling rate, which must be large enough to accurately reconstruct the system dynamics, yet not so large that the resulting measurement data exceeds available memory limits.


\section{Related Work}
\label{sec:related_work}
As described in \cite{Aguirre.16.07.2019}, the system identification process comprises five major steps: acquiring data, choosing a model class, determining its structure, estimating the parameters and, finally, validating the result. In the following literature review, we will focus on the different model classes, since most other identification steps depend on this crucial choice.

\textbf{Input-Output Models.} Popular approaches to identify nonlinear input-output behavior comprise for example radial basis functions \cite{vandenHof.1995}, Volterra-Series \cite{HASSOUNA2000947}, or the usage of an arbitrary nonlinear function in a nonlinear autoregressive model with exogenous inputs (NARX) \cite{Ma.2017}. Another possibility is the usage of established machine learning techniques, such as a decision tree \cite{Schoukens.2019} or a Gaussian process to model a state space representation of the system \cite{Sarkka.13.07.2019}. However, these methods are not designed for identifying an \ac{ODE}, but rather create a model for the input-output behavior and, thus, can identify a discrete-time difference equation at the most. Many control algorithms, however, require an \ac{ODE} instead of a discrete-time difference equation.

\textbf{Parameter Identification.} Identifying single parameters in an otherwise known \ac{ODE} is mostly solved by using various optimization algorithms. This task is sometimes referred to as \emph{inverse problem} and can be solved by gradient-based \cite{Rezk2022, MITRA20199} or gradient-free \cite{akman2018, Rudy2019} optimization algorithms. However, these algorithms often fail, if the \ac{ODE} exhibits chaotic, stiff, or highly nonlinear behavior, since it is necessary to solve the \ac{ODE} numerically in every optimization step. Machine learning algorithms, such as \acp{PINN} allow a parameter identification without calculating a numerical solution \cite{grimm2022, Nagel2022}.

\textbf{Identification of a Partly Known System.} In many cases the governing \ac{ODE} of a system is merely known to a certain extent, but it is not sufficiently precise for a real-world scenario. An example for this could be the modeling of a motor without considering the friction. It is possible to use a genetic algorithm to first estimate the parameters of the known system part and then extend it by a Gaussian process to model behavior, which goes beyond the scope of the governing equation~\cite{Caicedo.2017}. Another popular system identification method is \ac{SINDy} \cite{Brunton.2016}. It trains a matrix that comprises the coefficients of many nonlinear operators, which are used to construct an \ac{ODE}. The algorithm is well-established and has been improved to perform better under noise or in a control environment \cite{Fasel_2022}. A similar approach is given by the so-called \ac{EQL}, which replaces the standard activation functions of a neural network, like the hyperbolic tangent or a rectified linear unit, with other nonlinear functions that frequently appear in physical laws \cite{Sahoo.19.06.2018}. Both, the \ac{SINDy}-method and the \ac{EQL} can benefit from prior system knowledge, by pre-conditioning the used operators. 

\textbf{Identification of an Unknown System.} If only the system rank is known, the so-called subspace-based state-space estimation allows an identification of a linear state-space model by means of a QR-factorization as well as a singular value decomposition \cite{Viberg.2002}. Linear models can even be used in reinforcement learning \cite{Chen.2019}, but will suffer if the process features a strong nonlinear behavior. A more recent approach for identifying nonlinear models is the use of neural \acp{ODE} \cite{chen2018neural, Massaroli2020}, which allow a neural network to approximate an unknown \ac{ODE} by embedding it into a numerical \ac{ODE}-Solver and use the resulting solution to adjust the network weights based on training data. In \cite{Zhang.01.07.2022}, this method is combined with an \ac{EQL}, in order to allow the inclusion of symbolic elements in a neural \ac{ODE}-framework. However, this approach depends on the step width of the numerical \ac{ODE}-Solver and the identification quality degrades severely if the necessary step width becomes very small, hence, if the \ac{ODE}'s solution is stiff or highly oscillating. The inverse way of using integration instead of differentiation to fit an \ac{ODE} is presented in \cite{Mavkov2020} and is called \ac{INN}.

\section{Fundamentals}
\label{sec:fundamentals}
In this section, we present three fundamental methods, which our approach combines in order to identify an \ac{ODE} from noisy measurements: The \ac{EKBF} is a minimum variance estimator that allows estimating a system's states as precisely as possible. A \ac{PINN} allows approximating an \ac{ODE}'s solution without the usage of numerical solvers. Finally, an \ac{EQL} creates a special type of neural network, which uses operators that appear frequently in differential equations. Besides these three methods we briefly introduce model-based \ac{RL}.

\subsection{Extended Kalman-Bucy-Filter}
\label{sec:fundamentals_ekbf}
A Kalman filter allows estimating the system's state from noisy measurements and a given state space model. While the commonly known Kalman filter is limited to linear and discrete-time systems, the \ac{EKBF} performs state estimation for continuous-time and nonlinear systems \cite{Kalman.1961}. It does so by establishing two initial value problems, which describe the temporal evolution of the state's estimated mean value and covariance matrix. The mean value is given by
\begin{multline} \label{eq:kb_mean}
    \bm{\dot{\hat{x}}}(t) = \bm{f}\left(\bm{\hat{x}}(t),\bm{u}(t),\bm{0},t \right)  \\ + \bm{K}(t) \cdot \left( \bm{\overline{y}}(t) - \bm{g}(\bm{\hat{x}}(t), \bm{u}(t),\bm{0},t) \right)
\end{multline}
with a known initial value $\bm{\hat{x}}(t_0) = \bm{\hat{x}}_0$ and a Kalman gain
\begin{equation} \label{eq:kb_gain}
    \bm{K}(t) = \hat{\bm{P}}(t) \cdot \bm{\hat{C}}^\mathrm{T}(t) \cdot \bm{\hat{R}}^{-1}(t)
\end{equation}
as well as its covariance matrix
\begin{multline} \label{eq:kb_cov}
    \dot{\hat{\bm{P}}}(t) = \bm{\hat{A}}(t)\bm{\hat{P}}(t) + \bm{\hat{P}}(t)\bm{\hat{A}}(t)^\mathrm{T} \\ -  \bm{K}(t) \bm{\hat{C}}(t) \bm{\hat{P}}(t) + \bm{\hat{Q}}(t)
\end{multline}
with a known initial value $\bm{\hat{P}}(t_0) = \bm{\hat{P}}_0$ and linearized system matrices
\begin{equation}
\label{eq:linearizations}
\begin{aligned}
    \bm{\hat{A}}(t) &= \left. \frac{\partial \bm{f}(\bm{x},\bm{u}, \bm{w}, t)}{\partial \bm{x}(t)} \right|_{\wedge}~, &
    \bm{\hat{G}}(t) &=  \left. \frac{\partial \bm{f}(\bm{x},\bm{u}, \bm{w}, t)}{\partial \bm{w}(t)} \right|_{\wedge}~, \\
     \bm{\hat{C}}(t) &= \left. \frac{\partial \bm{g}(\bm{x},\bm{u}, \bm{v}, t)}{\partial \bm{x}(t)} \right|_{\wedge}~, &
    \bm{\hat{V}}(t) &= \left. \frac{\partial \bm{g}(\bm{x},\bm{u}, \bm{v}, t)}{\partial \bm{v}(t)} \right|_{\wedge}~.
\end{aligned}
\end{equation}
The $\wedge$-symbol denotes that the linearization is performed repeatedly for each new mean value $\bm{\hat{x}}(t)$. The covariance matrices of the process and measurement noise of the linearized system are given by means of
\begin{equation}
    \begin{split}
    \bm{\hat{Q}}(t) &= \bm{\hat{G}}(t) \cdot \bm{Q}(t) \cdot \bm{\hat{G}}^\mathrm{T}(t)~,\\
    \bm{\hat{R}}(t) &= \bm{\hat{V}}(t) \cdot \bm{R}(t) \cdot \bm{\hat{V}}^\mathrm{T}(t) ~,
    \end{split}
\end{equation}
respectively.

\subsection{Physics-Informed Neural Network}
\label{sec:fundamentals_pinn}
\acp{PINN} are a subclass of neural networks, which constrain the network's output to some known physical dynamics, which is usually given as a differential equation. They have been introduced by Raissi et al. \cite{raissi2017physics} for partial differential equations. In the following however, we define them in accordance to the state space representation of \eqref{eq:ss_representation}. Let $\mathcal{N}_\pi(t, \bm{u}|\bm W) \colon \mathbb{R}^{p+1} \to \mathbb{R}^n$ be a neural network, which maps the time $t$ and system input $\bm{u}$ to an estimated state vector $\tilde{\bm{x}}(t)$ by means of $\tilde{\bm{x}}(t) = \mathcal{N}_\pi(t,\bm u | \bm{W})$. Then, the neural network's weights $\bm{W}$ are trained by minimizing the loss
\begin{multline}
        J_\mathrm{PINN} = \sum_{i=1}^N\left( \frac{\mathrm{d}}{\mathrm{d}t} \tilde{\bm x}(t_i) - \bm{f}\left( \tilde{\bm{x}}(t_i),\bm{u}, t_i \right) \right)^2 \\ + \left(\tilde{\bm{x}}(t_0) - \bm{x}_0 \right)^2 ~.
        \label{eq:pinn_loss}
\end{multline}
Due to the time derivative of the neural network's output $\tilde{\bm{x}}(t)$ in \eqref{eq:pinn_loss}, it is necessary that $\mathcal{N}_{\pi}(t,\bm u|\bm W)$ contains only continuously differentiable activation functions, such as a hyperbolic tangent, which leads to the whole network being continuously differentiable. Note, however, that the original \ac{PINN} formulation does not consider any noise, neither process noise $\bm w$, nor measurement noise $\bm v$, in the system. Additionally, it assumes all states to be fully accessible and it does not consider a measurement function $\bm g(\cdot)$. 

\subsection{Equation Learner}
\label{sec:fundamentals_eql}
The \ac{EQL}, introduced by Sahoo et al. in \cite{Sahoo.19.06.2018}, learns equations that are suitable for extrapolation and control. The authors show that the method allows identifying a discrete-time state space model by creating a fully-connected neural network, which does not contain the commonly used activation functions such as hyperbolic tangent or rectified linear unit. Instead, they use continuously differentiable functions that appear frequently in governing equations of dynamical systems, such as sine, cosine, or multiplication. 
The last layer is composed of a division function. In order to keep the system continuously differentiable, the authors force the \ac{EQL} to output zero, if the denominator goes below a pre-defined threshold and add a penalty term forcing the system's output to avoid discontinuities. To obtain a difference equation with as few functional operators as possible, the authors also include $L_1$ and $L_0$ regularizations, which are activated in different phases of the training. However, the \ac{EQL} does not learn a continuous-time \ac{ODE}, but a discrete-time difference equation. Additionally, all system states are assumed to be directly accessible.
To address these limitations, we extend the \ac{EQL} in Sec.~\ref{sec:learning_odes}.

\subsection{Model-based Reinforcement Learning}
\ac{RL} is a subtopic of machine learning, which addresses the actions an agent has to take in an environment in order to maximize a reward. The interaction between agent and environment is defined by \iac{MDP}, which is a tuple $\left(\mathcal X, \mathcal U, \rho(\bm{x}_{k+1}|\bm{x}_k,\bm{u}_k), R(\bm{x}_k, \bm{x}_{k+1}, \bm{u}_k), \bm{x}_0\right)$ with the set of states $\mathcal X$ and the set of system inputs $\mathcal U$ \cite{wang2019benchmarking}. $\rho(\bm{x}_{k+1}|\bm{x}_k,\bm{u}_k)$ describes a probability distribution to reach the successor state $\bm{x}_{k+1}$, provided a current state $\bm x_k$ and input $\bm u_k$. $R(\bm{x}_k, \bm{x}_{k+1},\bm{u}_k)$ is the associated reward to this transition. Both functions are not known in the \ac{RL} context and are learned implicitly. 

The target is to find a control strategy (or \emph{policy}) $K(\cdot): \mathbb{R}^n \rightarrow  \mathbb{R}^p$ with $\bm{u}_k = K(\bm{x}_k)$, which maximizes the expected reward $V = \mathbb{E} \{ \sum_{i=k}^{\tau} \gamma^i\cdot R(\bm x_i, \bm u_i) \}$ with a discount factor $\gamma \in \left(0,1\right)$. In model-based \ac{RL}, the discrete-time transition function is explicitly trained by using data, which is generated by the agent exploring the environment. The routine is described as follows: First, the agent performs random actions in the environment and, thus, gathers data. Afterwards, a dynamic model is trained and becomes utilized to find a control function $K(\cdot)$, which is often realized as \ac{MPC}~\cite{Zanon.2021}. After applying the control function to the system, new data is acquired, which is used to improve the model's accuracy and, thus, allows updating the control law. In this work, we use an identified \ac{ODE} to find an optimal control function.

Note that \ac{RL} typically considers purely discrete-time problems and models. However, our approach allows the identification of a continuous-time model, which avoids the adverse effects of temporal discretization during the design of a controller.

\section{Learning ODEs}
\label{sec:learning_odes}
\begin{figure*}[tb]
\centering
    \begin{subfigure}[t]{0.52\textwidth}
    \centering
    \resizebox{0.8\linewidth}{!}{
\begin{tikzpicture}
        \draw (2.25,1.15) circle [radius=0.35];
        \node at (2.25,1.15) {$1$};
        \draw[-stealth] (2.6,1.15) -- (4.2,-0.3);

        \draw (2.25,0.15) circle [radius=0.35];
		\node at (2.25,0.15) {$z_1$};
        \draw[-stealth] (2.6,0.15) -- (4.2,-0.3);

        \draw (2.25,-1.85) circle [radius=0.35];
		\node at (2.25,-1.85) {$z_O$};
        \draw[-stealth] (2.6,-1.85) -- (4.2,-0.3);          	    

   	    \node at (3.4,1.0) {$\bm{W}^{(1)}$};
   	    
         \draw (5,2) circle [radius=0.6];
         \draw (5,1.4) -- (5,2.6);
         \node at (4.7,2) {$\sum$};
         \node at (5.3,1.95) {\small{$\mathrm{op}_1$}};
         
         \draw (5,0.5) circle [radius=0.6];
    		 \draw (5,-0.1) -- (5,1.1);
    		 \node at (4.7,0.5) {$\sum$};
    		 \node at (5.3,0.45) {\small{$\mathrm{op}_2$}};
    		 
    		 \node at (5,-0.3) {$\mathbf{\vdots}$};
    		 
          \draw (5,-1.25) circle [radius=0.6];
    		 \draw (5,-1.85) -- (5,-0.65);
    		 \node at (4.7,-1.25) {$\sum$};
    		 \node at (5.31,-1.3) {\small{$\mathrm{op}_K$}};
    		 
             \draw (5,-2.75) circle [radius=0.6];
    		 \node at (5,-2.75)  {$1$};
    		 
    		 \draw (7,-0.3) circle [radius=0.6];
    		 \draw (7,-0.9) -- (7,0.3);
    		 \node at (6.7,-0.3) {$\sum$};
    		 \node at (7.3,-0.3) {$(\cdot)$};
    		 
    		 \draw[->] (5.6,2) -- (6.55,0.1);
    		 \draw[->] (5.6,0.5) -- (6.4,-0.3);		 
    		 \draw[->] (5.6,-1.25) -- (6.4,-0.3);
    		 \draw[->] (5.6,-2.75) -- (6.55,-0.7);
    		 
    		 \node at (6.5,1.2) {$\bm{W}^{(2)}$};

    		 \draw (4.2,2.8) -- (7.8,2.8) -- (7.8,-3.55) -- (4.2,-3.55) -- cycle;
    		 \draw (7.8,2.7) -- (7.9,2.7) -- (7.9,-3.65) -- (4.3,-3.65) -- (4.3,-3.55); 
 		 \draw (7.9,2.6) -- (8.0,2.6) -- (8.0,-3.75) -- (4.4,-3.75) -- (4.4,-3.65); 

 		 \draw[->](7.85,2.85) -- (8.15,2.55);
 		 \node at (8.25,2.85) {$\times I$};
 		 
 		 \draw (9.25,-0.675) circle [radius=0.7];
 		 \node at (9.25,-0.675) {$\prod_{i=1}^{I}$};
 		 
 		 \draw[-stealth] (7.6,-0.3) -- (8.65,-0.3);
 		 \draw[-stealth] (7.8,-0.675) -- (8.55,-0.675);
 		 \draw[-stealth] (7.9,-1.05) -- (8.65,-1.05);
 		 
 		 \draw (10.75,-0.675) circle [radius=0.35];
 		 \node at (10.75,-0.675) {$o$}; 
 		 \draw[-stealth] (9.95,-0.675) -- (10.4,-0.675);		 
	\end{tikzpicture}}
    \caption{Operator-neuron}
    \label{fig:ode_neuron} 
    \end{subfigure}%
    \qquad
    \begin{subfigure}[t]{0.34\textwidth}
    \centering
    \resizebox{0.8\linewidth}{!}{
\begin{tikzpicture}
		\draw (0,10) circle [radius=0.35];
		\node at (0,10) {$\bm{x}$};
		
		\draw (1,10) circle [radius=0.35];
		\node at (1,10) {$\bm{u}$};
		
		\draw (2,10) circle [radius=0.35];
		\node at (2,10) {$\bm{w}$};		
		
		\draw (3,10) circle [radius=0.35];
		\node at (3,10) {$t$};
		
		\draw[-stealth] (0,9.65) -- (0,9);
		\draw[-stealth] (0,9.65) -- (2.5,9);
		\draw[-stealth] (1,9.65) -- (0,9);
		\draw[-stealth] (1,9.65) -- (2.5,9);
		\draw[-stealth] (2,9.65) -- (0,9);
		\draw[-stealth] (2,9.65) -- (2.5,9);
		\draw[-stealth] (3,9.65) -- (0,9);
		\draw[-stealth] (3,9.65) -- (2.5,9);

		\draw (-1.5,9) -- (1,9) -- (1,8.5) -- (-1.5,8.5) -- cycle;
		\node[] at (-0.25,8.72) {\footnotesize{Op-neuron $1,1$}};
		\node at (1.25,8.72) {\footnotesize{$\mathbf{\mydots}$}};
		\draw (1.5,9) -- (4,9) -- (4,8.5) -- (1.5,8.5) -- cycle;
		\node at (2.75,8.72) {\footnotesize{Op-neuron $1,N_1$}};
		
		\draw (-1.5,7.5) -- (1,7.5) -- (1,7) -- (-1.5,7) -- cycle;
		\node at (-0.25,7.22) {\footnotesize{Op-neuron $M,1$}};
	 	\node at (1.25,7.22) {\footnotesize{$\mydots$}};
		\draw (1.5,7.5) -- (4,7.5) -- (4,7) -- (1.5,7) -- cycle;
		\node at (2.75,7.22) {\footnotesize{Op-neuron $M,N_M$}};
		
		\draw[-stealth] (0,8.5) -- (0,7.5);
		\draw[-stealth] (0,8.5) -- (2.5,7.5);
		\draw[-stealth] (2.5,8.5) -- (2.5,7.5);
		\draw[-stealth] (2.5,8.5) -- (0,7.5);
		\node[fill=white] at (1.25,8.1) {$\mathbf{\vdots}$};

        \node[] at (1.25,6.75) {\footnotesize{$\bm{w}^{(3)},\bm{w}^{(4)}$}};
		
		\draw (1.25,6.0) circle [radius=0.5];
		\draw (0.75,6.0) -- (1.75,6.0);
		\node at (1.25,6.25) {$\sum$};
		\node at (1.25,5.75) {$\frac{~\cdot~}{~\cdot~}$};
		
		\draw[-stealth] (-0.15,7) -- (0.95,6.4);
		\draw[-stealth] (2.65,7) -- (1.55,6.4);	
		
		\draw (1.25,4.75) circle [radius=0.35];
		\node at (1.25,4.75) {$\dot{x}$};
		\draw[-stealth] (1.25,5.5) -- (1.25,5.1);
			 
	\end{tikzpicture}}
    \caption{\ac{ODE}-Network}
    \label{fig:eql_network}
    \end{subfigure}
    \caption{Sketches of the proposed architecture of a single operator-neuron (left) and of the whole \ac{ODE}-Network (right).}
\end{figure*}
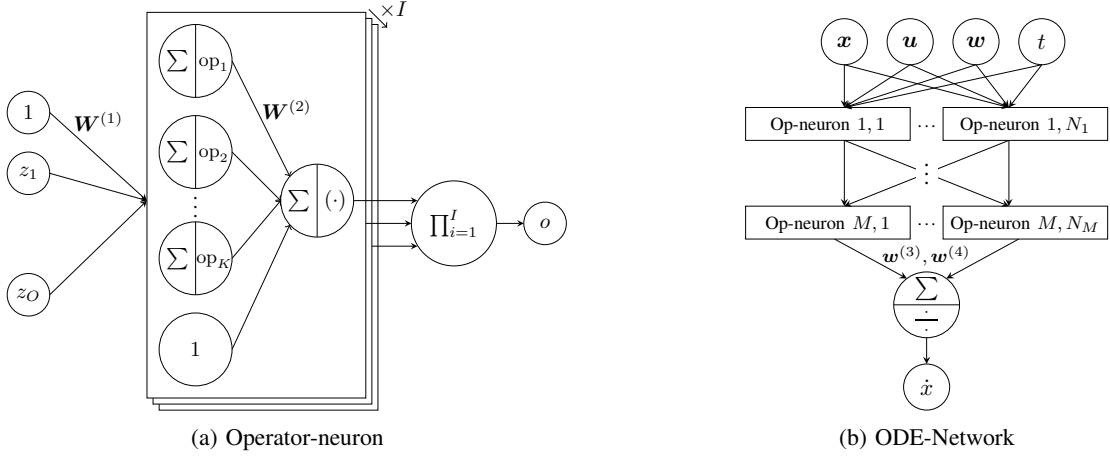

In the following, we introduce our \ac{ODE}-Learner framework, which is the main contribution of this paper. Therefore, we build a special \ac{EQL}-architecture and include it in a \ac{PINN}-framework to identify the correct \ac{ODE}. To handle noisy data and modeling errors, we combine the approach with \iac{EKBF}. Finally, we integrate several regularization terms in order to acquire an \ac{ODE}, which is as compact as possible.

\subsection{ODE-Network}
\label{sec:ode_network}

The overall target of our method is to identify the nonlinear ODE $\dot{\bm{x}} = \bm{f}(\bm{x},\bm{u},\bm{w},t)$ of \eqref{eq:ss_representation}. Therefore, we create a special neural network $\mathcal{N}_\zeta$, which we call \emph{ODE-Network}, to fit the unknown ground truth $\bm{f}(\cdot)$ as correctly as possible. The main difference between a conventional \ac{MLP} and the \ac{ODE}-Network is the usage of activation functions, which appear frequently in dynamic systems, such as trigonometric functions or polynomial elements. This idea has first been introduced as \ac{EQL} in \cite{Sahoo.19.06.2018} (cf. Sec.~\ref{sec:fundamentals_eql}). In many systems, the true dynamics arises from the multiplication of several nonlinear operators. To achieve this behavior in the original implementation of the \ac{EQL}, a very deep network has to be trained. Instead, we create a special \ac{EQL}-Neuron, which we call \emph{operator-neuron}. It allows building a rather shallow network, at the expense of an increased width, i.e., an increasing neuron count. Afterwards, we combine several operator-neurons to an \ac{ODE}-Network. This aims at creating an \ac{ODE}, which is as compact as possible. 

Let the operator-neuron's input be denoted by $\bm{z} = \left[1, z_1,\ldots,z_O \right]^\mathrm{T} \in \mathbb{R}^{O+1}$ with a bias in the beginning and its output by $o \in \mathbb{R}$. Additionally, we define an operator set $\mathcal O = \{\mathrm{op}_{1}(\cdot),\ldots,\mathrm{op}_{K}(\cdot)\}$, which lists $K$ different operators. Note that every operator is required to be continuously differentiable. Finally, the operator-neuron exhibits two weight matrices $\bm{W}^{(1)} \in \mathbb{R}^{I \times K \times (O+1)}$ and $\bm{W}^{(2)} \in \mathbb{R}^{I \times (K+1)}$. One operator-neuron then performs a forward pass by calculating
\begin{equation}
\label{eq:ode_neuron}
    o = \prod_{i=1}^{I} \left( \bm W_{i,K+1}^{(2)} + \sum_{k=1}^{K} \bm W_{i,k} ^{(2)}\cdot \mathrm{op}_{k}\left(\sum_{l=1}^{O+1} \bm W_{i,k,l}^{(1)} \cdot z_l\right) \right) ~.
\end{equation}
The input to the neuron is weighted by the first weighting matrix $\bm{W}^{(1)}$ and summed. Afterwards, it is transformed by the set of $K$ operators. The result is then again weighted by the second weighting matrix $\bm{W}^{(2)}$ and a bias is added in order to allow a constant activation, if necessary. The operator-neuron performs these calculations $I$ times in parallel and multiplies the scalar results, respectively. Fig.~\ref{fig:ode_neuron} shows a sketch of the calculations that are performed in \eqref{eq:ode_neuron}. The final \ac{ODE}-Network is created by combining multiple operator-neurons in a similar way to a fully-connected neural network with $M$ layers and $N_j$, $j=1,\ldots, M$, neurons per layer. By using operator-neurons of the proposed structure, a complicated \ac{ODE} can be built even with a shallow \ac{ODE}-Network, because of the possibility to directly include multiplication elements. The last layer is composed of two weight vectors $\bm{w}^{(3)} = \left[ w^{(3)}_0, \ldots, w^{(3)}_{N_M} \right]^\mathrm{T} \in \mathbb{R}^{N_m+1}$, $\bm{w}^{(4)} = \left[ w^{(4)}_0, \ldots, w^{(4)}_{N_M} \right]^\mathrm{T} \in \mathbb{R}^{N_m+1}$, which we use to calculate the ratio
\begin{equation}
\dot{x} = \begin{cases}
\frac{\bm{o}^\mathrm{T} \cdot \bm{w}^{(3)}} {\bm{o}^\mathrm{T} \cdot \bm{w}^{(4)}} & \mathrm{if} ~ \bm{o}^\mathrm{T} \cdot \bm{w}^{(4)} > \delta \\
0 & \mathrm{otherwise}
\end{cases}~,
\end{equation}
with the output of the preceding layer $\bm{o} = \left[1, o_1, \ldots, o_{N_M} \right]^\mathrm{T}$ and a hyperparameter $\delta > 0$. This ensures that there is no division by zero. Additionally, we include a penalty term to the loss function, which leads to a higher loss if the denominator is below $\delta$ (cf. Sec.~\ref{sec:ode-learner_kbinn}). This idea has first been introduced in \cite{Sahoo.19.06.2018}. Fig.~\ref{fig:eql_network} shows a sketch of the \ac{ODE}-Network's architecture.
\begin{figure*}[!t]
    \centering
    \resizebox{0.7\linewidth}{!}{
\begin{tikzpicture}
		\node at (1.5,10.5) {$t$};
		\draw[-stealth] (1.5,10.35) -- (1.5,10);	
		\draw[blue, line width=0.9pt] (-0.3,10) -- (3.3,10) -- (3.3, 9) -- (-0.3,9) -- cycle;
		\node at (1.5,9.5) {Mean-network $\mathcal{N}_\xi$};
		
		\draw[-stealth] (1.5,9) -- (1.5, 8.8);
		\node at (1.5,8.6) {$\bm{\xi}$};
		
		\draw (0,8.3) -- (0.5,8.3) -- (0.5,7.6) -- (0,7.6) -- cycle;
		\node at (0.25,7.95) {$\frac{\mathrm{d}}{\mathrm{d}t}$};
		\draw[-stealth, blue, line width=0.9pt] (1.3,8.6) -- (0.25,8.6) -- (0.25,8.3);
		\draw[-stealth, blue, line width=0.9pt] (0.25,7.6) -- (0.25,7.2);		
		\node at (0.25,7) {$\dot{\bm{\xi}}_\mathrm{AD}$};
		\draw[-stealth, blue, line width=0.9pt] (0.25,6.8) -- (0.25,6.5) -- (1,6.5);

		\draw[-stealth, blue, line width=0.9pt] (1.7,8.6) -- (2.5,8.6) -- (2.5,7.5) -- (3.7,7.5);
		\draw[-stealth, blue, line width=0.9pt] (1.7,8.6) -- (3.5,8.6) -- (3.5,10.5) -- (4.75,10.5) -- (4.75,10);
		\draw[-stealth, blue, line width=0.9pt] (1.7,8.6) -- (2.5,8.6) -- (2.5,8.0) -- (9.25,8.0) -- (9.25,6.75);
		
		\draw (1,6.75) -- (2,6.75) -- (2,6.25) -- (1,6.25) -- cycle;
		\node at (1.5,6.5) {$L_2$};
		
		\node at (5.5,10.5) {$t$};
		\draw[-stealth] (5.5,10.35) -- (5.5,10);	
		\node at (6.25,10.5) {$u$};
		\draw[-stealth] (6.25,10.35) -- (6.25,10);
		
		\draw[mygreen, line width=0.9pt] (3.7,10) -- (7.3,10) -- (7.3, 9) -- (3.7,9) -- cycle;
		\node at (5.5,9.5) {State-network $\mathcal{N}_\zeta$};
		\draw[-stealth,mygreen, line width=0.9pt] (5.5,9) -- (5.5, 8.8);
		\node at (5.5,8.6) {$\dot{\bm{\xi}}$};
		
		\draw[-stealth,mygreen, line width=0.9pt] (5.5,8.35) -- (5.5,8.25) -- (1.2,8.25) -- (1.2,6.75);
		\draw[-stealth,mygreen, line width=0.9pt] (5.5,8.35) -- (5.5,8.25) -- (9.75,8.25) -- (9.75,6.75);
		
		\node at (9.5,10.5) {$t$};
		\draw[-stealth] (9.5,10.35) -- (9.5,10);	
		
		\draw[violet, line width=0.9pt] (7.7,10) -- (11.3,10) -- (11.3, 9) -- (7.7,9) -- cycle;
		\node at (9.5,9.5) {Covariance-network $\mathcal{N}_\psi$};
		\draw[-stealth,violet, line width=0.9pt] (9.5,9) -- (9.5, 8.8);
		\node at (9.5,8.6) {$\bm{\psi}$};

		\draw (11,8.3) -- (10.5,8.3) -- (10.5,7.6) -- (11,7.6) -- cycle;
		\node at (10.75,7.95) {$\frac{\mathrm{d}}{\mathrm{d}t}$};
		\draw[-stealth,violet, line width=0.9pt] (9.65,8.6) -- (10.75,8.6) -- (10.75,8.3);
		\draw[-stealth,violet, line width=0.9pt] (10.75,7.6) -- (10.75,7.2);		
		\node at (10.75,7) {$\dot{\bm{\psi}}_\mathrm{AD}$};
		\draw[-stealth,violet, line width=0.9pt] (10.75,6.8) -- (10.75,6.5) -- (10,6.5);		
		
		\draw[-stealth,violet, line width=0.9pt] (9.35,8.6) -- (8.6,8.6) -- (8.6,7.5) -- (7.3,7.5);
		\draw[-stealth,violet, line width=0.9pt] (9.35, 8.6) -- (8.6,8.6) -- (8.6,8.125) -- (1.8,8.125) -- (1.8,6.75);
		
		\draw (9,6.75) -- (10,6.75) -- (10,6.25) -- (9,6.25) -- cycle;
		\node at (9.5,6.5) {$L_3$};		
		
		\draw[cyan] (3.7,7.9) -- (7.3,7.9) -- (7.3, 7) -- (3.7,7) -- cycle;
		\node[] at (5.5,7.45) {Output-network $\mathcal{N}_\eta$};
		
		\draw[-stealth,cyan, line width=0.9pt] (5.5,7) -- (5.5,6.5) -- (2,6.5);
		\draw[-stealth,cyan, line width=0.9pt] (5.5,7) -- (5.5,6.5) -- (9,6.5);
		\draw[-stealth,cyan, line width=0.9pt] (5.5,7) -- (5.5,6.15);
		
		\draw (4.8,6.15) -- (6.2,6.15) -- (6.2,5.65) -- (4.8,5.65) -- cycle;
		\node at (5.5,5.9) {$L_1$};
		
		\node at (1.5,5.9) {$\overline{\bm{y}}(t)$};	
		\draw[-stealth](1.9,5.9) -- (4.8,5.9);
		\draw[-stealth](1.5,6.1) -- (1.5,6.25);		
	\end{tikzpicture}}
    \caption{Sketch of the ODE-Learner concept, describing the interactions of the first three loss functions. The different colors indicate the influence of a network to the respective loss.}
    \label{fig:main_concept}
\end{figure*}
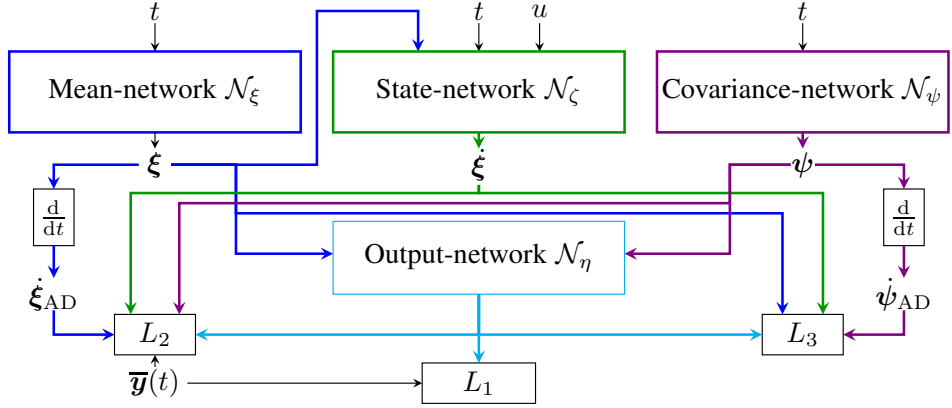

The usage of the \ac{ODE}-Network features multiple benefits over an \ac{MLP}. The most important one is the possibility to include prior existing system knowledge by the choice of an operator set. E.g., if we want to model an inverse pendulum system, it makes sense to include trigonometric operators, such as $\mathcal O = \{\sin(\cdot), \cos(\cdot)\}$. Using an \ac{MLP} usually discards this prior knowledge. Depending on our system understanding, it is possible to go even further. In case of situation (i), we do not use \iac{ODE}-Network at all, but simply use the known \ac{ODE} and only train the unknown parameters. Situation (ii) is solved by pre-conditioning the weight matrices of the \ac{ODE}-Network, based on the existing system knowledge.
During the following training, the pre-conditioned \ac{ODE}-Network is adapted so it suits the provided data. ``Adapted'' means that the given ODE can be extended by new equation elements and that the initial \ac{ODE}'s parameters can become readjusted. Hence, we use the initial ODE as baseline and improve it continuously during the training. Situation (iii) demands the most training efforts, since, similar to solving a machine learning problem, the architecture needs to be determined by trial-and-error and the operators are set to the best of the user's knowledge. Nevertheless, even without knowing the system dynamics, it is still often possible to make good guesses on the operator set. Another benefit of the \ac{ODE}-Network over an \ac{MLP} is the improved extrapolation behavior, which has already been investigated in \cite{Sahoo.19.06.2018} for the \ac{EQL}. This is especially important for using the \ac{ODE}-Network as a model in a model-based \ac{RL} framework in order to acquire a high reward after as few iterations as possible.

The \ac{ODE}-Network cannot be trained in a supervised fashion, because it maps the state $\bm x$, the input $\bm u$, and the process noise $\bm w$ on the state's temporal derivative $\dot{\bm{x}}$. However, $\dot{\bm{x}}$ is not available in the measurement data and it also cannot be precisely estimated by calculating the difference quotient, especially if only noisy measurements are provided. To circumvent this problem, we use an \ac{ODE}-Learner framework, which is introduced in the following subsection.

\subsection{The ODE-Learner Framework}
\label{sec:ode-learner_kbinn}
The \ac{ODE}-Learner framework allows fitting an \ac{ODE} by providing noisy data without simplifying it to a discrete-time difference equation by combining \iac{PINN}-approach with \iac{EKBF}. The approach to solve \iac{EKBF} in a physics-informed way has first been introduced in \cite{Nagel2022} as \ac{KBINN}, with the aim on estimating unknown real-valued parameters in an otherwise known state-space model. In this section, we extend the \ac{KBINN} by including the \ac{ODE}-Network in order to identify a continuous-time, nonlinear system with noisy measurements. Here, the \ac{ODE}-Network estimates the dynamics in the form of a state-space representation. However, in order to fit a state-space representation to data, the \ac{ODE} of the state needs to be solved, which is done by the \ac{KBINN}.

We implement four neural networks in total as shown in Fig.~\ref{fig:main_concept}. The \emph{mean-network} $\mathcal{N}_\xi$ is a standard fully-connected neural network that approximates the state's mean vector by means of $\hat{\bm{x}}(t) \approx \bm{\xi}(t) = \mathcal{N}_\xi(t)$ with the network's output $\bm \xi(t)$. The \emph{covariance-network} $\mathcal{N}_\psi(t)$ is also a fully-connected neural network, which approximate the state's covariance matrix by means of $\hat{\bm{P}}(t) \approx \bm{\psi}(t) = \mathcal{N}_\psi(t)$ with the network output $\bm \psi(t)$. The \emph{state-network} $\mathcal{N}_\zeta$ is an \ac{ODE}-Network and aims at approximating the \ac{ODE} $\bm f(\cdot)$ in \eqref{eq:ss_representation} by using the state's estimated mean value according to $\dot{\bm x} = \mathcal{N}_\zeta(\bm{x},\bm{u},\bm{w},t)$. Finally, the \emph{output-network} $\mathcal{N}_\eta$ is also an \ac{ODE}-network and approximates the system's measurement equation $\bm g(\cdot)$ by calculating $\bm y(t) = \mathcal{N}_\eta(\bm{x},\bm{u},\bm{v},t)$. 

The idea is summarized as follows: We train the networks in a \ac{PINN}-framework, which is constrained by the differential equations of an \ac{EKBF}'s mean value and covariance matrix that we introduced in \eqref{eq:kb_mean} and \eqref{eq:kb_cov}. This approach allows including measurement noise and process noise. Additionally, we do not need every state to be directly accessible. Hence, we train the \ac{ODE}-Learner by minimizing a loss function that comprises four terms according to
\begin{equation} \label{eq:objective_function}
    L = \frac{1}{N}\sum_{i=1}^N \left( \alpha_1 L_{1,i} + \alpha_2 L_{2,i} + \alpha_3 L_{3,i} + \alpha_4 L_{4,i}\right)~,
\end{equation}
with weights $\alpha_j > 0$, $j \in \left\{1,2,3,4 \right\}$ to adjust the influence of the single components. The first term $L_{1,i}$ is a maximum-likelihood function that keeps our identified system's output close to the measurements $\overline{\bm{y}}(t_i)$ according to
\begin{equation}
    L_{1, i} = -\sum_{j=1}^q \log \left( \mathcal{L}_j\left(\bm{\xi}(t_i),\bm{\psi}(t_i) ; \overline{y}_{j}(t_i)\right)  \right) ~,
\end{equation}
with the scalar output values $\overline y_j(t_i)$, $j = 1,\ldots,q$ of the measurement vector $\overline{\bm{y}}(t_i)$\,. Since the noise is assumed to be Gaussian, $\mathcal{L}_j$ describes the normal probability distribution function by means of 
\begin{equation}
   \mathcal{L}_j(\cdot) = \frac{1}{\sqrt{2\pi \sigma_j^2(t_i)}} \exp \left\{\frac{\left(\overline{y}_j(t_i)-\mu_j(t_i)\right)^2}{2\sigma_j^2(t_i)}\right\}
\end{equation}
The mean value $\mu_j(t_i)$ and variance $\sigma_j^2(t_i)$ are the first two stochastic moments of the estimated state, when propagated through the output network by means of 
\begin{equation}
    \begin{split}
        \mu_j(t_i) &= \mathrm{mean}\left( \eta_j(t_i) \right)~\mathrm{and}\\
        \sigma_j^2(t_i) &= \mathrm{var}\left( \eta_j(t_i) \right)
    \end{split}
\end{equation}
Here, $\eta_j(t_i)$ is acquired by propagating the estimated mean value $\bm \xi$ and the estimated covariance matrix $\bm \psi$ through the output-network $\bm \eta(t_i) = \mathcal{N}_\eta(\bm \xi, \bm \psi; \bm u, \bm v, t_i)$. Then, the output's mean value and variance can be both calculated in closed-form if the function $\mathcal{N}_\eta(\cdot)$ is (piece-wise) linear, trigonometric, or polynomial \cite{Huber.2015}. If the mean and variance cannot be calculated in closed-form, an approximate calculation by sampling or a numerical integration is required. Hence, it is advantageous to constrain the output-network to the mentioned cases.


The second term $L_{2,i}$ is defined by means of
\begin{equation}
        L_{2,i} = \lnorm\bm{\xi}(t_0) - \bm{x}_0 \rnorm_2 + \lnorm \dot{\bm{\xi}}_\mathrm{AD}(t_i) - \bm{\Xi}(\bm{\xi},t_i) \rnorm_2
\end{equation}
with
\begin{multline}
    \label{eq:Xi}
    \bm{\Xi}(\bm{\xi},t_i) = \mathcal{N}_\zeta\left( \bm{\xi}(t_i),\bm{u},\bm{0}, t_i\right) \\ + \bm{K}(t_i) \cdot \left(\overline{\bm{y}}(t_i) - \mathcal{N}_\eta(\bm{\xi}(t_i),\bm{u},\bm{0},t_i) \right)~ .
\end{multline}
and the Kalman gain $\bm{K}(t_i) = \bm{\psi}(t_i) \cdot \bm{\hat{C}}^\mathrm{T}(t_i) \cdot \bm{\hat{R}}^{-1}(t_i)$. $\dot{\bm{\xi}}_\mathrm{AD}$ denotes the temporal derivative of $\bm{\xi}(t)$, acquired by automatic differentiation. Note that \eqref{eq:Xi} describes the mean's temporal evolution in an \ac{EKBF}, which we introduced in \eqref{eq:kb_mean}. In its essence, $L_{2,i}$ ensures that the state-network $\mathcal{N}_\zeta$ approximates the derivative with respect to time of the mean-network $\mathcal{N}_\xi$, while reducing the influence of measurement or process noise.

The third term $L_{3,i}$ is defined by means of
\begin{equation}
    L_{3,i} = \lnorm \bm{\psi}(t_0) - \hat{\bm{P}}_0 \rnorm_\mathrm{F} + \lnorm \dot{\bm{\psi}}_\mathrm{AD}(t_i) - \bm{\Psi}(\bm{\psi},t_i) \rnorm_\mathrm{F}
\end{equation}
and
\begin{multline}\label{eq:Psi}
     \bm{\Psi}(\bm{\psi},t_i) = \bm{\hat{A}}(t_i)\bm{\psi}(t_i) + \bm{\psi}(t_i)\bm{\hat{A}}(t_i)^\mathrm{T} \\ - \bm{K}(t_i) \bm{\hat{C}}(t_i) \bm{\psi}(t_i) + \bm{\hat{Q}}(t_i)~.
\end{multline}
The matrices in \eqref{eq:Psi} are acquired by linearization in a similar way to \eqref{eq:linearizations} by replacing $\bm f(\cdot)$ with $\mathcal{N}_\zeta(\cdot)$, $\bm g(\cdot)$ with $\mathcal{N}_\eta(\cdot)$ and $\hat{\bm x}$ with $\bm{\xi}$. $\dot{\bm \psi}_\mathrm{AD}$ is the temporal derivative of $\bm \psi(t)$, acquired by automatic differentiation.
Hence, the third loss enforces the temporal evolution of the estimated covariance matrix $\bm{\psi}(t)$ to be consistent with the estimated state-space models $\mathcal{N}_{\zeta}$ and~$\mathcal{N}_{\eta}$.

Summarized, $L_{1,i}$ keeps the estimated system's output close to the measurement data, while $L_{2,i}$ and $L_{3,i}$ force the estimated system to be compatible to the \ac{EKBF}'s mean and covariance matrix. The fourth loss $L_{4,i}$ is formed from regularization terms, which are described in the following. We want to enforce the network to only use the operators, which are deemed necessary. At the same time, we do not want to punish large coefficient values, since they might be necessary for a successful identification. Thus, the first regularization term is given by 
\begin{equation}
    \label{eq:ode_regularization}
    R_0(w) = \frac{a_1}{1+\exp\{-a_2\cdot \left|w\right| + a_3\}} + a_4\cdot \left|w\right|
\end{equation}
with constants $a_1,\ldots, a_4$ and a scalar weight value $w$. The regularization performs a very low punishment, if the weight value is close to zero and increases heavily afterwards. For larger values, the increase in regularization is marginal but present, which avoids vanishing gradients. 

As we already mentioned in Sec.~\ref{sec:ode_network}, it is necessary to punish an approach of the \ac{ODE}-Network's denominator to the pole. Thus, we define another regularization term 
\begin{equation}
    R_1(\bm o, \bm w) = \max(0, \delta - \bm{o}^\mathrm{T} \cdot \bm{w}^{(4)}) ~,
\end{equation}
which is similar to the method used in \cite{Sahoo.19.06.2018}. The loss $L_{4,i}$ from \eqref{eq:objective_function} is created from
\begin{equation}
    L_{4,i} = \alpha_{4,1} R_0(w) + \alpha_{4,2} R_1(w)
\end{equation}
with additional weights $\alpha_{4,1},\alpha_{4,2} > 0$\,.

\section{Validation}
\label{sec:validation}
We validate our method on three benchmarks. First, we focus on pure system identification by training \iac{ODE}-Learner and solve the resulting \ac{ODE} numerically in order to compare the predicted outcome to the ground truth. Afterwards, we embed our \ac{ODE}-Learner in a model-based \ac{RL}-framework in order to reach a target state by using as little data as possible. As it is mentioned in Sec.~\ref{sec:ode_network}, the \ac{ODE}-Network can be substituted with a standard \ac{MLP}. However, this does neither allow the inclusion of prior knowledge, nor can \ac{MLP}s extrapolate very well. Nevertheless, in the following experiments, we compare the results of using an \ac{ODE}-Network with a standard \ac{MLP}. Additionally, we use state of the art algorithms that allow identifying a continuous-time \ac{ODE} to compare it to our \ac{ODE}-Learner framework. To be more precise, we choose SINDy, which is well-established in the system identification community, Mathwork's neural state-space, which is a Matlab-implementation of a neural \ac{ODE} and, finally, the \ac{INN}, as a counter-part to the neural \ac{ODE} that does not rely on differentiation. We already introduced all three algorithms in Sec.~\ref{sec:related_work}. To obtain an implementation, we use the publicly available GitHub-repositories for \ac{SINDy} \cite{desilva2020} and the \ac{INN} \cite{Mavkov2020}, respectively. The neural state-space model is available in Matlab's system identification toolbox at version R2022b or later. We want to emphasize that we have chosen the hyperparameters of the methods to the best of our knowledge and belief. Furthermore, it is important to note that the results may vary significantly with different hyperparameters. In the following, we give an overview of our experiments.

\subsection{Duffing Oscillator}
\begin{figure}[tb]
    \centering
    \def\doPath{./data/duffing_oscillator/rmse_list.csv}

\begin{tikzpicture}
\begin{axis}[
boxplot/draw direction=y,
xtick={1,2,3,4,5},
x tick label style={
    align=center
},
xticklabels={{ODE-\\Network}, MLP, SINDy, Matlab, INN},
width=\linewidth,
height=0.6\linewidth,
ymajorgrids=true,
grid style=dashed,
ylabel={$e_\mathrm{RMSE}$},
ymode=log,
legend to name={legend_do_ct},
legend cell align={left},
ymax=3
]

\addplot[boxplot, color=red] table[y=CODELearner, col sep=comma]{\doPath};
\addlegendentry{ODE-Learner}
\addplot[boxplot, color=blue] table[y=MLP, col sep=comma]{\doPath};
\addplot[boxplot, color=blue] table[y=SINDY, col sep=comma]{\doPath};
\addlegendentry{SotA}
\addplot[boxplot, color=blue] table[y=MATLAB, col sep=comma]{\doPath};
\addplot[boxplot, color=blue] table[y=INN, col sep=comma]{\doPath};

\end{axis}
\end{tikzpicture} 
    \caption{Root mean squared error after identifying the Duffing oscillator with ten different parameter scenarios. Dots indicate outliers.}
    \label{fig:system_identification_errors}
\end{figure}
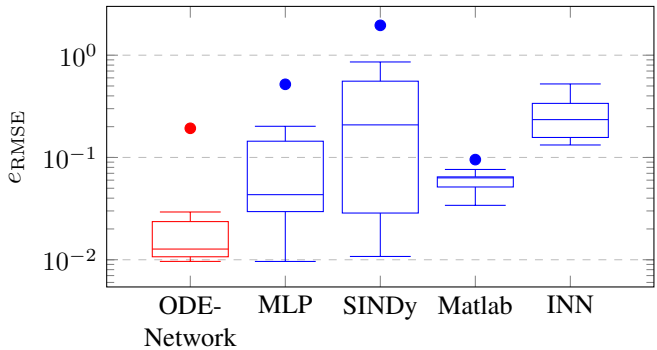
The Duffing oscillator is a 2\textsuperscript{nd}-order benchmark system, which is very popular in the system identification community \cite{Goharoodi2018, Wigren2013}. It contains nonlinear operators and has a chaotic behavior, which means it is crucial to obtain a model that is able to generalize well from the provided data. Its state-space equation is given by means of
\begin{equation} \label{eq:ss_duffing_oscillator}
\begin{split}
    \dot{x}_1(t) &= x_2(t) \\
    \dot{x}_2(t) &= b_1 x_2(t) + b_2 x_1(t) + b_3 x_1(t)^3 - b_4 \cos(\omega_0 t) \\
    y(t) &= x_1(t) + v(t) ~,
\end{split}
\end{equation}
with parameters $b_1,\ldots,b_4\in \mathbb{R}$, which define the system behavior, and the frequency of the input signal $\omega_0 \in \mathbb{R}$. We assume that we do not know the state-space equation in the identification process, apart from the operators, and only use \eqref{eq:ss_duffing_oscillator} to synthesize data. Note, however, that we exacerbate the usual problem by only making $x_1(t)$ measurable after adding some zero-mean Gaussian-distributed noise $v(t)$. 

\begin{table}[b]
    \centering
    \caption{Root mean squared errors of the system identification tasks for the Duffing oscillator and the Cascaded Tank.}
    \begin{tabular}{l|c c}
     \textbf{Method} & \textbf{Duffing oscillator} & \textbf{Cascaded Tank} \\
     \midrule
     \textbf{\ac{ODE}-Learner}  & $0.03 \pm 0.06$ &   $0.61$ \\
     \textbf{\ac{MLP}}  & $0.11 \pm 0.16$ &   $1.23$ \\
     \textbf{\ac{SINDy}} & $0.43 \pm 0.61$ & $1.53$ \\
     \textbf{Matlab} & $0.06 \pm 0.02$ & $1.26$ \\
     \textbf{\ac{INN}} & $0.27 \pm 0.13$ & $0.81$ $(0.41)$\footnotesize{\textsuperscript{1}} \\
    \end{tabular}\\
    \flushleft{\footnotesize{\textsuperscript{1}The $e_\mathrm{RMSE}$ value of 0.41 was reported in \cite{Mavkov2020}.}}
    \label{tab:identification_results}
\end{table}
To prove the effectiveness of the \ac{ODE}-Learner, we build ten systems by sampling $b_1,\ldots,b_4$, $\omega_0$ and the initial states $x_1(0)$, $x_2(0)$ from a uniform distribution $\mathcal{U}(0,1)$. Afterwards, we simulate measurement data by using Euler's method to solve \eqref{eq:ss_duffing_oscillator} numerically over a period of $t_\mathrm{end}=\SI{48}{s}$ with a sampling time of $\Delta t = \SI{0.02}{s}$ and add zero-mean Gaussian noise according to $v(t) \sim \mathcal N(0,0.01^{2})$. We use the first $\SI{40}{s}$ of measurement data as a training set and the last $\SI{8}{s}$ to validate our model. We identify a 2\textsuperscript{nd}-order \ac{ODE}, which corresponds to situation (iii) according to Section~\ref{sec:introduction}, i.e., the entire ODE needs to be identified. The Operator-neuron allows a cubic calculation. Afterwards, we use Euler's method to create a simulated trajectory of the states and feed them through the output equation to obtain the estimated system output $\hat{y}(t_i)$. The quality of the identification is then measured by calculating the root mean squared error $e_\mathrm{RMSE}$ between the identification $\hat{y}(t_i)$ and the ground truth $y(t_i)$. 

The results of all identification runs are shown in Fig.~\ref{fig:system_identification_errors} as a boxplot. The \ac{ODE}-Learner using an \ac{ODE}-Network enables a more precise identification than the other methods. Substituting the \ac{ODE}-Network with an \ac{MLP} leads to a slightly worse fit. In one out of the ten identification runs, we observe an outlier, which is caused by a gradient being stuck in a local minimum during the training. This could be corrected by restarting the training with new weights, however we kept the results in Fig.~\ref{fig:system_identification_errors} unchanged. Additionally, we want to emphasize that the Matlab method requires all states to be accessible. Thus, we simplified the problem for Matlab and used both states. Additionally, Table~\ref{tab:identification_results} shows the mean value and standard deviation of the obtained identification errors.


\subsection{Cascaded Tanks}
In this paragraph we test the effectiveness of our method on a real-world system identification benchmark \cite{Schoukens2020}. It is composed of two vertically arranged water tanks and a water reservoir, which is placed below. Fig.~\ref{fig:cascaded_tank_sketch} features a sketch of the system.

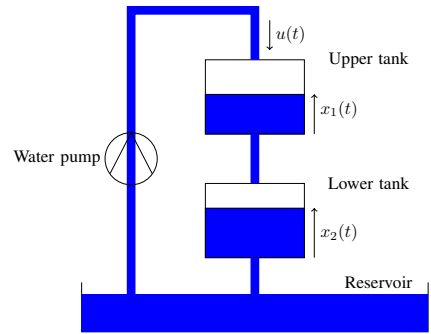
\begin{figure}[t]
    \centering
    \resizebox{0.65\linewidth}{!}{
    \begin{tikzpicture}
	\draw (0.5,0.5) rectangle (2.5,2);
	\draw[fill=blue, fill opacity=0.5] (0.5,0.5) rectangle (2.5,1.3);
	\node[] at (3.8,2) {Upper tank};
	\draw[->] (2.7,0.5) -- (2.7,1.3);
	\node[anchor=west] at (2.7,1) {$x_1(t)$};
	
	\draw (0.5,-0.5) rectangle (2.5,-2);
	\draw[fill=blue, fill opacity=0.5] (0.5,-2) rectangle (2.5,-1);
	\node[] at (3.8,-0.5) {Lower tank};
	\draw[->] (2.7,-2) -- (2.7,-1);
	\node[anchor=west] at (2.7,-1.5) {$x_2(t)$};

	\draw (-2,-2.5) -- (-2,-3.5) -- (5,-3.5) -- (5,-2.5);	
	\path[fill opacity=0.5, fill=blue] (-2,-2.75) rectangle (5,-3.5);
	\node[] at (4,-2.5) {Reservoir};
	
	\draw[line width=5 pt, color=blue] (-1,-3) -- (-1,3) -- (1.5,3) -- (1.5,2);
	\draw[line width=5 pt, color=blue] (1.5,0.5) -- (1.5,-0.5);
	\draw[line width=5 pt, color=blue] (1.5,-2) -- (1.5,-3);
	\draw[->] (1.8,2.8) -- (1.8,2.2);
	\node[anchor=west] at (1.8,2.5) {$u(t)$};
	
	\filldraw[fill=none] (-1,0) circle (15 pt);
	\draw (-1.425,-0.3) -- (-1,0.525) -- (-0.575,-0.3);
	\node[anchor=east] at (-1.5,0) {Water pump};	
\end{tikzpicture}}
    \caption{Sketch of the cascaded tank system.}
    \label{fig:cascaded_tank_sketch}
\end{figure}
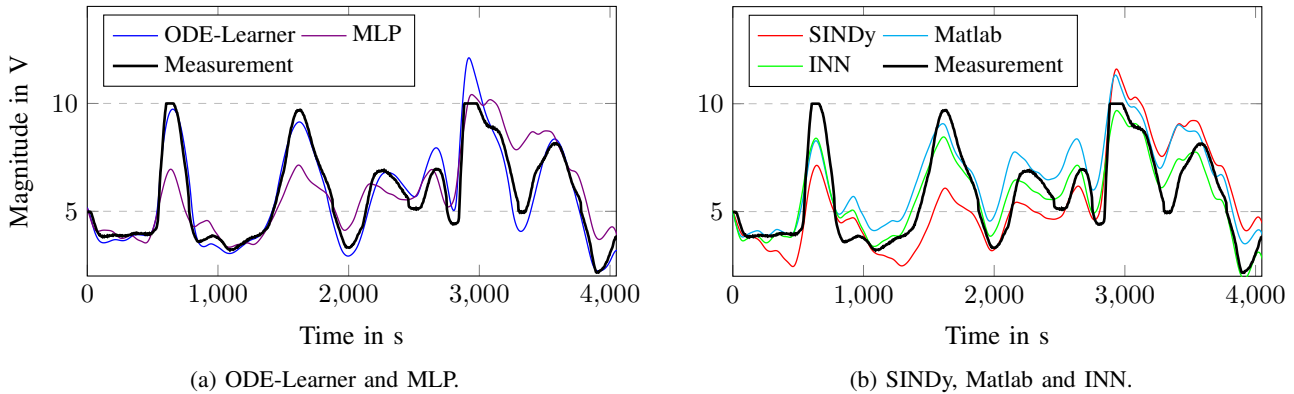
\begin{figure*}[t!]
    \centering
    \def\dataPath{./data/cascaded_tank/cascaded_tank_outcome.csv}
\def\dataOdeLearner{./data/cascaded_tank/ct_simulation.csv}

\begin{subfigure}[t]{0.475\textwidth}
    \centering
    \begin{tikzpicture}
    \begin{axis}[
        ylabel={Magnitude in V},
        xlabel={Time in s},
        xmin=0, xmax=4048,
        ymin=2, ymax=14.5,
        legend pos=north west,
        ymajorgrids=true,
        grid style=dashed,
        width=\linewidth,
        height=0.6\linewidth,
        legend cell align={left},
        legend style={font=\small, legend columns=2}
        ]
        \addplot[color=blue, line width=0.5pt] table[x=t, y=y_val_kbinn, col sep=comma] {\dataPath}; \addlegendentry{ODE-Learner}
        \addplot[color=violet, line width=0.5pt] table[x=t, y=y_val_mlp, col sep=comma] {\dataPath}; \addlegendentry{MLP}
        \addplot[color=black, line width=1pt] table[x=t, y=y_val, col sep=comma]{\dataPath};\addlegendentry{Measurement} 
    \end{axis}
    \end{tikzpicture}
    \caption{ODE-Learner and MLP.}
     \label{fig:mean and std of net14}
    \end{subfigure}%
~
    \begin{subfigure}[t]{0.475\textwidth}  
    \centering 
    \begin{tikzpicture}
    \begin{axis}[
        xmin=0, xmax=4048,
        ymin=2, ymax=14.5,
        legend pos=north west,
        xlabel={Time in s},
        ymajorgrids=true,
        grid style=dashed,
        width=\linewidth,
        height=0.6\linewidth,
        legend cell align={left},
        legend style={font=\small, legend columns=2}
        ]
        \addplot[color=red, line width=0.5pt] table[x=t, y=y_val_sindy, col sep=comma] {\dataPath}; \addlegendentry{SINDy}
        \addplot[color=cyan, line width=0.5pt] table[x=t, y=y_val_matlab, col sep=comma] {\dataPath};  \addlegendentry{Matlab}
        \addplot[color=green, line width=0.5pt] table[x=t, y=y_val_INN, col sep=comma] {\dataPath}; \addlegendentry{INN}
        \addplot[color=black, line width=1pt] table[x=t, y=y_val, col sep=comma]{\dataPath}; \addlegendentry{Measurement}
    \end{axis}
    \end{tikzpicture}
    \caption{SINDy, Matlab and INN.}  
    \label{fig:mean and std of net24}
    \end{subfigure}%
	\caption{Plots of the simulated height for the cascaded tank problem. (a) shows the result of our identified ODE, obtained by an \ac{ODE}-Network and an \ac{MLP}, compared to the original test data, while (b) shows state of the art system identification methods.}
	\label{fig:cascaded_tank_simulation}
\end{figure*}
A water pump transports the water from the reservoir into the upper tank. Each tank features a small opening at its bottom, allowing water to flow from the upper tank to the lower one and, finally, from the lower tank into the reservoir. However, both tanks do not have a lid. If the water pump's power is too high, both tanks can overflow. This behavior is nonlinear and stochastic, since water from the upper tank can flow partly into the lower tank or into the reservoir. The benchmark authors provide an \ac{ODE}, which describes the state-space of the water flow by using Bernoulli's equation according to
\begin{equation} \label{eq:ss_cascaded_tank}
    \begin{split}
        \dot{x}_1(t) &= -k_1\sqrt{x_1(t)} + k_4 u(t) + w_1(t) \\
        \dot{x}_2(t) &= k_2\sqrt{x_1(t)} - k_3\sqrt{x_2(t)} + w_2(t) \\
        y(t) &= x_2(t) + v(t) ~,
    \end{split}
\end{equation}
with $x_1(t)$ and $x_2(t)$ denoting the water level in the upper and lower tank, respectively. $u(t)$ denotes the system input and $k_1,\ldots,k_4$ are unknown system parameters.  The benchmark authors use a D/A-converter as system input, which controls the pump's power by setting a voltage. The system's output is the water level of the lower tank, which is measured by a capacitive water level sensor. The measurement values are, again, voltage values, this time from an A/D-converter. Neither sensor nor water pump dynamics are provided, which are, together with the overflow behavior and the initial state values $\bm{x}_0$, subject to identification. Hence, this use case corresponds to situation (ii) according to Section~\ref{sec:introduction}, i.e., some segments as well as all parameters of the ODE need to be identified. The benchmark authors provide a training and a test data set with 1,024 instances, respectively, which are each sampled at $\Delta t = \SI{4}{s}$. We train an \ac{ODE}-Learner by providing a pre-conditioned \ac{ODE}-Network according to the dynamics, given in \eqref{eq:ss_cascaded_tank}. We extend the network by nine operator-neurons to model the unknown dynamics. The output network used only performs a linear transformation, since we assume the capacitive water level sensor to exhibit linear behavior.

The $e_\mathrm{RMSE}$ of the identified systems are shown in Table~\ref{tab:identification_results}. Note that we can perform this validation only once, since it is a real-world example and we do not have multiple data sets available. Fig.~\ref{fig:cascaded_tank_simulation} shows the plots and error values of the simulated outputs and the provided validation data for each investigated method. It is visible that we identified \iac{ODE} that shows the most similar behavior to the original one. The \ac{INN} is very close to ours. We hypothesize that this is caused by the low sampling rate of only $\SI{4}{s}$, which is too coarse for differentiation-based methods, but better suits methods that rely on integration. The authors of the \ac{INN}-method even reached a root-mean-squared error of $0.41$ with another hyperparameter setting. Replacing the \ac{ODE}-Network by an MLP leads to results being similar to \ac{SINDy} and Matlab.

\subsection{Inverted Pendulum on a Cart}
\begin{figure*}[t]
    \centering
    \def\cartpoleSimulationOne{data/cartpole/cartpole_simulation_1.csv}
\def\cartpoleSimulationTwo{data/cartpole/cartpole_simulation_2.csv}
\def\cartpoleSimulationThree{data/cartpole/cartpole_simulation_3.csv}

\begin{subfigure}[t]{0.45\linewidth}
    \centering
    \input{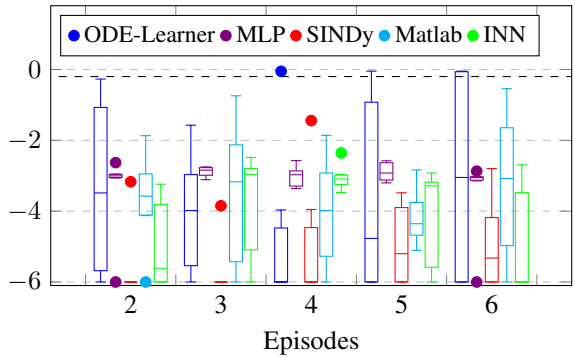}
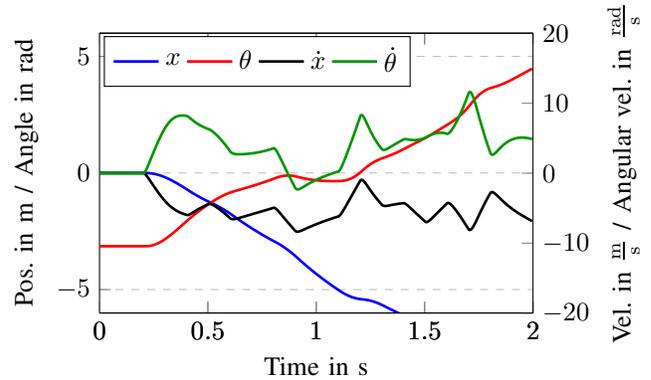
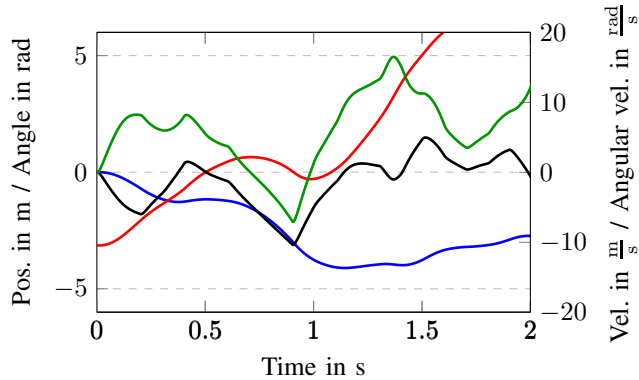
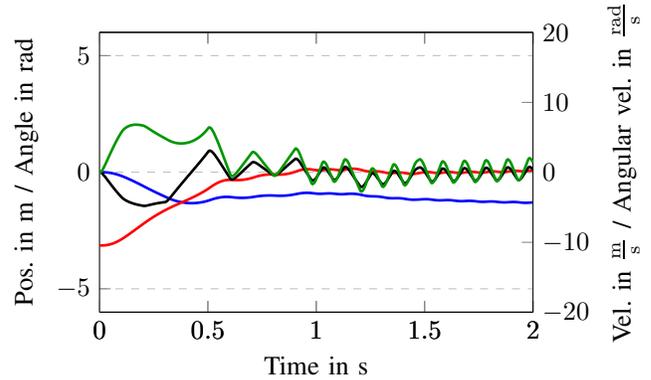
    \caption{Achieved rewards of the inverted pendulum. The dashed black line indicates a successful swing-up.}
    \label{fig:cartpole_car_boxplot}
\end{subfigure}
~
\begin{subfigure}[t]{0.45\linewidth}
\centering
\begin{tikzpicture}
	\begin{axis}[
        axis y line*=left,
		xlabel={Time in s},
		ylabel={Pos. in m / Angle in rad},
		xmin=0, xmax=2,
		ymin=-6, ymax=6,
		ymajorgrids=true,
		grid style=dashed,
        ylabel near ticks,
		width=0.9\linewidth,
		height=0.65\linewidth,
		]
		\addplot[color=blue, line width=1pt] table[x=t, y=x0, col sep=comma] {\cartpoleSimulationOne};\label{plot_x_cp}

		\addplot[color=red, line width=1pt] table[x=t, y=x2, col sep=comma] {\cartpoleSimulationOne};\label{plot_phi}
	\end{axis}
    \begin{axis}[
    axis y line*=right,
    ylabel={Vel. in $\mathrm{\frac{m}{s}}$ / Angular vel. in $\mathrm{\frac{rad}{s}}$},
    ymin=-20, ymax=20,
    xmin=0, xmax=2,
    ylabel near ticks,
    width=0.9\linewidth,
	height=0.65\linewidth, 
    legend style={at={(0.01,0.99)},anchor=north west,legend columns=-1}
    ]
    \addlegendimage{/pgfplots/refstyle=plot_x_cp}\addlegendentry{$x$}
    \addlegendimage{/pgfplots/refstyle=plot_phi}\addlegendentry{$\theta$}
    
    \addplot[color=black, line width=1pt] table[x=t, y=x1, col sep=comma] {\cartpoleSimulationOne}; \addlegendentry{$\dot{x}$}

    \addplot[color=mygreen, line width=1pt] table[x=t, y=x3, col sep=comma] {\cartpoleSimulationOne}; \addlegendentry{$\dot{\theta}$}
    \end{axis}
\end{tikzpicture}
\caption{Episode 1: Random actions}
\label{fig:cartpole_episode_1}
\end{subfigure}
\\
\begin{subfigure}{0.45\linewidth}
\centering
\begin{tikzpicture}
	\begin{axis}[
        axis y line*=left,
		xlabel={Time in s},
		ylabel={Pos. in m / Angle in rad},
		xmin=0, xmax=2,
		ymin=-6, ymax=6,
		ymajorgrids=true,
		grid style=dashed,
        ylabel near ticks,
        width=0.9\linewidth,
		height=0.65\linewidth,
		]
		\addplot[color=blue, line width=1pt] table[x=t, y=x0, col sep=comma] {\cartpoleSimulationTwo};

		\addplot[color=red, line width=1pt] table[x=t, y=x2, col sep=comma] {\cartpoleSimulationTwo};

	\end{axis}
    \begin{axis}[
    axis y line*=right, 
    ylabel={Vel. in $\mathrm{\frac{m}{s}}$ / Angular vel. in $\mathrm{\frac{rad}{s}}$},
    ymin=-20, ymax=20,
    xmin=0, xmax=2,
    ylabel near ticks,
    width=0.9\linewidth,
    height=0.65\linewidth,
    ]
    \addplot[color=black, line width=1pt] table[x=t, y=x1, col sep=comma] {\cartpoleSimulationTwo};

    \addplot[color=mygreen, line width=1pt] table[x=t, y=x3, col sep=comma] {\cartpoleSimulationTwo};
    \end{axis}
\end{tikzpicture}
\caption{Episode 2: Pure MPC, because the switching criterion has not been fulfilled.}
\label{fig:cartpole_episode_2}
\end{subfigure}
~
\begin{subfigure}{0.45\linewidth}
\centering
\begin{tikzpicture}
	\begin{axis}[
        axis y line*=left,
		xlabel={Time in s},
		ylabel={Pos. in m / Angle in rad},
		xmin=0, xmax=2,
		ymin=-6, ymax=6,
		ymajorgrids=true,
		grid style=dashed,
        ylabel near ticks,
		width=0.9\linewidth,
		height=0.65\linewidth,
		]
		\addplot[color=blue, line width=1pt] table[x=t, y=x0, col sep=comma] {\cartpoleSimulationThree}; 

		\addplot[color=red, line width=1pt] table[x=t, y=x2, col sep=comma] {\cartpoleSimulationThree};

	\end{axis}
    \begin{axis}[
    axis y line*=right,  
    ylabel={Vel. in $\mathrm{\frac{m}{s}}$ / Angular vel. in $\mathrm{\frac{rad}{s}}$},
    ymin=-20, ymax=20,
    xmin=0, xmax=2,
    ylabel near ticks,
    width=0.9\linewidth,
	height=0.65\linewidth,
    ]

    \addplot[color=black, line width=1pt] table[x=t, y=x1, col sep=comma] {\cartpoleSimulationThree};

    \addplot[color=mygreen, line width=1pt] table[x=t, y=x3, col sep=comma] {\cartpoleSimulationThree};
    \end{axis}
\end{tikzpicture}
\caption{Episode 3: MPC until $t=\SI{0.83}{s}$. Afterwards an LQR controls the input}
\label{fig:cartpole_episode_3}
\end{subfigure}
    \caption{Achieved results for the inverse pendulum on a cart. The first plot shows a box plot of obtained reward after different episodes for the investigated methods. The next three plots show the trajectory of the controlled cart for one showcase, where we reached a swing-up at the third episode.}
    \label{fig:cartpole_simulation}
\end{figure*}
The inverted pendulum on a cart is a frequently used benchmark for controllers. A pendulum is mounted on a movable cart, which has one degree of freedom. The control target is to swing up the pendulum from its lower equilibrium point and keep it stable in an upright position. The system has a rank of $n=4$ and a state vector $\bm{x} = \begin{bmatrix}
x & \dot{x} & \theta & \dot{\theta}\end{bmatrix}^\mathrm{T}$ with the cart position $x$, the cart velocity $\dot x$, the pendulum angle $\theta$, and the angular velocity $\dot \theta$. The use case corresponds to situation (iii) according to Section~\ref{sec:introduction}. The pendulum is in an upright position, when $\theta = 0$. 

In the literature, the control job is often split up into the swing-up and the stabilization task \cite{Mori1976}. Thus, we use \iac{MPC}-algorithm to swing up the pendulum. When it reaches its target state, \iac{LQR} keeps it stable in an upright position. We define the reward function by means of $R = \frac{1}{d}\sum_{i=N-d}^N -|\theta(t_i)|$ with the pendulum angle $\theta(t_i)$ at time step $t_i$ and a time window length $d\in \mathbb N$. We consider the pendulum stable, if $R > -0.2$\,. This allows a very small oscillation of less than $\pm 11.5^\circ$ around the target position to be counted as successful swing-up. The initial state of the system is $\bm{x}_0 = \begin{bmatrix} 0 & 0 & -\pi & 0 \end{bmatrix}^\mathrm{T}$.  We perform the following routine: First, we simulate training data by moving a cart with a random input signal, which is capped at $\SI{-25}{N} < u < \SI{25}{N}$. Afterwards, we train a first \ac{ODE} by using the randomly gathered data. We then use this identified \ac{ODE} in an \ac{MPC}-framework to perform an L-BFGS-B-optimization to calculate the optimal input values, which we apply to the inverse pendulum system. Each episode comprises a data sample of $t_\mathrm{end} = \SI{2}{s}$ with a sampling rate of $\Delta t = \SI{4}{ms}$. The \ac{MPC} solves an optimization problem of minimizing $J_\mathrm{MPC} = -R(\bm{x})$ over a prediction horizon of $n_\mathrm{ph} = \SI{1}{s}$. We switch from the \ac{MPC} to an \ac{LQR}, as soon as the angle is close to the target position for a short period. To be precise, $|\theta| < \pi/6$ during $95\,\%$ of the last $\SI{0.2}{s}$. When this switching criterion is reached, we use the identified \ac{ODE} and calculate an \ac{LQR}. 

Fig.~\ref{fig:cartpole_car_boxplot} shows the obtained reward for different episodes after six identification runs with different inital input values. In three out of six cases, we could swing up the pendulum and stabilize it after six episodes. In the remaining three cases, we did not observe a swing-up in the first six episodes. With more than six episodes, we obtained a large amount of training data, which becomes increasingly difficult to handle with the \ac{ODE}-Learner, which is why we refrained from performing further episodes and considered these attempts as failed. We did not observer a swing-up with the other methods. Note that we capped the negative reward at $R_\mathrm{min} = -6$.
Fig.~\ref{fig:cartpole_episode_1} to \ref{fig:cartpole_episode_3} show the trajectories of a successful swing-up after three episodes, which we observed in another test run as a showcase.

\section{Discussion}
\label{sec:discussion}
In this section, we want to list the strengths and limitations of our method. As we showed in Sec.~\ref{sec:validation}, our method reliably allows identifying \iac{ODE}, even when noisy data is provided. Additionally, it is possible to include human knowledge, which is not the case for methods like a neural \ac{ODE}, the \ac{INN} or an \ac{MLP}. The prediction error remains reasonably small and our method is also applicable in a model-based \ac{RL} framework. Furthermore, it is possible to extract a differential equation with a better root mean squared error than established methods. 

However, there are some drawbacks. Firstly, the training time is rather high. We used an Nvidia A100 GPU with training times between $\SI{45}{min}$ for the duffing oscillator and $\SI{36}{h}$ for the complete run of the inverted pendulum task. This is due to the need to linearize the state-network in order to obtain the state matrix $\bm A$ in \eqref{eq:Psi}, which requires multiple differentiation steps. If we do not manage a successful run in the first few episodes of a model-based \ac{RL}-problem, the accumulated data grows to such an extent that the training time becomes unreasonably high, which is what happened in the inverted pendulum task. All other methods require significantly less training time. Additionally, our method has a large number of hyperparameters. Especially the architecture of all four networks has a major impact on the result. The only way to find the correct network architecture is by trial-and-error or by utilizing a neural architecture search \cite{ren2021comprehensive}. As is visible in Fig.~\ref{fig:system_identification_errors}, the \ac{ODE}-Learner sometimes shows an outlier. We observe this behavior in approximately one in ten cases. We suspect that the reason for this lies in the backpropagation algorithm, which can get stuck in a local minimum. If we restart the algorithm with new weighting matrices, the outlier disappears. Nevertheless, we are still performing better than the state of the art in terms of data efficiency, since there is no method available that allows swinging up a pendulum after less than $\SI{6}{s}$ of measurement data. Another drawback is the non-interpretability of the identified \acp{ODE}. Despite experimenting a lot with increasing the weight of the introduced regularization terms, we were not able to identify \acp{ODE}, which are compact enough to interpret the mathematical model and to visually check, whether the model is correct. Finally, despite identifying a continuous-time differential equation, the control is performed by an \ac{MPC}-algorithm, which discretizes the identified \ac{ODE} afterwards. Using a continuous-time control law might leverage the advantages of continuous-time identification further.

\section{Conclusion and Outlook}
\label{sec:conclusion}
We presented a new method to identify \acp{ODE} from noisy measurement data that allows human knowledge to be incorporated and that does not demand every state to be directly accessible. We showed that the identification works very precisely and proved that it is possible to embed it into a model-based \ac{RL} framework. For future work, we plan to optimize the control strategy, in order to use the trained \acp{ODE} more efficiently and reduce the data usage even further. Additionally, we are working on training a nonlinear control function in parallel to the identified \ac{ODE}.


\bibliographystyle{plain}        
\bibliography{bibliography}           

\end{document}